\documentstyle[twocolumn,psbox,myletter]{jpsj}

\def\runtitle{
Spin Fluctuation-Induced Superconductivity in $\kappa$-BEDT-TTF Compounds
}
\def\runauthor{Hisashi {\sc Kondo} and T\^{o}ru {\sc Moriya}}

\title
{
Spin Fluctuation-Induced Superconductivity \\
in $\kappa$-BEDT-TTF Compounds
}

\author
{ 
Hisashi {\sc Kondo} and T\^{o}ru {\sc Moriya}
}

\inst
{
Department of Physics, 
Faculty of Science and Technology, Science University of Tokyo, Noda 278-8510
}

\recdate
{\today}

\abst
{
    Spin fluctuation-induced superconductivity in quasi-two dimensional organic
compounds, $\kappa$-BEDT-TTF salts, is investigated within a fluctuation exchange 
(FLEX) approximation using a half-filled Hubbard model with a right-angled
isosceles triangular lattice (transfer matrices $-{\mit \tau}$,
$-{\mit \tau}^{\prime}$), extending a previous work above $T_{\rm c}$.
An energy gap of ${\rm A}_2$ or $(x^2-y^2)$-type develops with decreasing 
temperature below $T_{\rm c}$ more rapidly than in the BCS model. The calculated 
dynamical
susceptibilities enough below $T_{\rm c}$ show sharp resonance peaks like those
in certain cuprates superconductors. The calculated nuclear spin-lattice
relaxation rate $1/T_1$ shows a $T^3$ behavior below $T_{\rm c}$ in accordance
with experiment. Estimated values of $1/T_1$ are roughly consistent with 
experimental results. A prediction is made for the doping concentration 
dependence of $T_{\rm c}$ and the antiferromagnetic and superconductive 
instability points are calculated in the $U/{\mit \tau}$ vs. 
${\mit \tau}^{\prime}/{\mit \tau}$ plane.
}

\kword
{
organic superconductor, spin fluctuation, Hubbard model, triangular lattice,
resonance peak, nuclear spin-lattice relaxation rate
}

\begin{document}
\sloppy
\maketitle

%%%\newpage

Recent theoretical investigations indicate that the superconductivity in 
quasi-two dimensional organic compounds $\kappa$-$({\rm BEDT}$-${\rm TTF})_2X$  
[where $X = {\rm Cu} \{ {\rm N(CN)_2} \} X^{\prime}$, $X^{\prime} = {\rm Cl}$, 
${\rm Br}$] can be explained in terms of the spin fluctuation 
mechanism.~\cite{rf:1,rf:2,rf:3,rf:4,rf:5} The calculated values for 
the transition temperature $T_{\rm c}$ as well as the d-wave character of 
the order parameter seem to be consistent with the existing experimental 
results.~\cite{rf:6,rf:22,rf:7,rf:23,rf:24,rf:8}

In a previous paper (hereafter referred to as I) we have reported the results 
of calculations for $T_{\rm c}$ and the normal state properties of these 
compounds by using a Hubbard model consisting of dimer orbitals.~\cite{rf:1} 
The purpose of the present communication is to report on the results of our 
extended calculations mainly to include the superconducting state. 

The model and the approach are the same as those discussed in I. We use a 
half-filled Hubbard model with a right-angled isosceles triangular lattice 
consisting of the anti-bonding dimer orbitals with the inter-dimer transfer 
integrals $-{\mit \tau}$ and $-{\mit \tau}^{\prime}$ as shown in Fig. 1.
The estimated values for the transfer integrals are 
${\mit \tau} \sim 0.07 {\rm eV}$ and 
${\mit \tau}^{\prime}/{\mit \tau} \sim 0.8$.~\cite{rf:9,rf:10} 
We use the fluctuation exchange (FLEX) approximation where the dynamical 
susceptibilities are calculated within the renormalized random phase 
approximation (RRPA) and the normal and the anomalous self-energies are 
approximated by the simplest ones including a spin fluctuation propagator 
without vertex corrections. This approximation has been successful in many 
previous investigations on high-$T_{\rm c}$ cuprates, 
etc.~\cite{rf:11,rf:12,rf:13,rf:14,rf:15,rf:16,rf:17,rf:18} and 
recent studies on the vertex corrections seem to indicate that its effects 
will not seriously modify the FLEX results.~\cite{rf:19}
\begin{figure}[t]
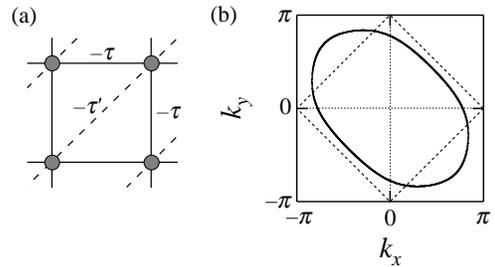

  \begin{center}
    \psbox[size=0.55#1]{fig1.eps}
  \end{center}
  \caption{
  (a) The model unit cell and the transfer integrals.
  (b) Unperturbed Fermi surface for ${\mit \tau}^{\prime} / {\mit \tau} = 0.8$.
  Dashed lines show the antiferromagnetic zone boundary.
  }
\end{figure}

Before reporting on the results below $T_{\rm c}$ we show in Fig. 2 
the calculated phase boundaries, in the $U/{\mit \tau}$ vs. 
${\mit \tau}^{\prime}/{\mit \tau}$ plane, between the paramagnetic metallic  
(PM) and the antiferromagnetic (AF) phases and between PM and the 
superconductive (SC) phases. The former is obtained from the points where the 
value of $1/{\mit \chi}_{\mibs Q}$ extrapolated to $T = 0$ vanishes. The latter 
is obtained from the points of vanishing $T_{\rm c}$ calculated by extrapolating 
the $T_{\rm c}/{\mit \tau}$ vs. $U/{\mit \tau}$ plots, 
shown in the inset of Fig. 2, using the Pad\'{e} approximants.~\cite{rf:27} 
This result, being consistent with the conjectures in I, shows 
that the AF and SC instabilities compete with each other and the latter wins 
for the values of ${\mit \tau}^{\prime}/{\mit \tau}$ larger than $\sim 0.3$. 
In other words, the d-wave superconductivity appears near the antiferromagnetic 
instability only when the electronic structure is favorable; an 
antiferromagnetic instability is not necessarily accompanied by a d-wave 
superconductivity. With increasing $U/{\mit \tau}$ from a superconducting 
instability point we expect a phase transition to an antiferromagnetic state. 
The phase boundary between the superconducting and antiferromagnetic phases 
must be obtained by comparing the free energies of the two phases, since the 
phase transition is naturally of the first order. This task is left for future 
studies.
\begin{figure}[t]
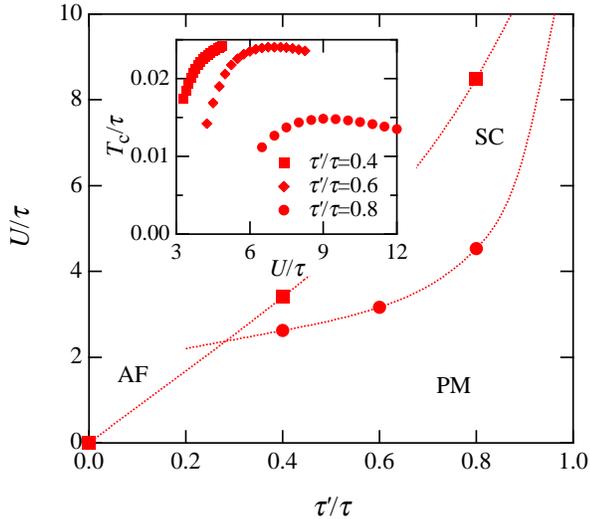

  \begin{center}
    \psbox[size=0.65#1]{fig2.eps}
  \end{center}
  \caption{
     Instability points of the paramagnetic state against 
     superconductivity and antiferromagnetism. Inset shows the results 
     of an improved calculation for $T_{\rm c}/{\mit \tau}$ vs. $U/{\mit \tau}$.
  }
\end{figure}

Figure 3 shows the calculated values of $T_{\rm c}$ for 
${\mit \tau}^{\prime}/{\mit \tau}=0.8$ for various values of $n$, the number 
of electrons per site, and $U/{\mit \tau}$. These results may be regarded 
as a prediction for the doping concentration-dependence of $T_{\rm c}$ in 
$\kappa$-$({\rm BEDT}$-${\rm TTF})_2X$.  
\begin{figure}[t]
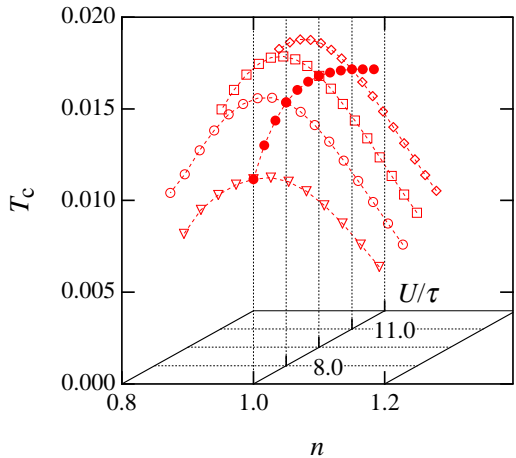

  \begin{center}
    \psbox[size=0.65#1]{fig3.eps}
  \end{center}
  \caption{
     Electron number (doping concentration) dependence of 
     $T_{\rm c}/{\mit \tau}$ for ${\mit \tau}^{\prime} / {\mit \tau} = 0.8$.
  }
\end{figure}

Now we discuss the properties in the superconducting state. We first show 
in Figs. 4(a) and 4(b) the anisotropy and the temperature dependences of 
the gap function 
${\mit \Delta}_{\mibs k}= {\rm Re} 
                \{ {\mit \Delta}({\mib k},{\mit \Delta}_{\mibs k}) \}$, 
respectively, where the gap function is defined by
\begin{eqnarray}
     {\mit \Delta}({\mib k},{\mit \omega}) &=&
     \frac{{\mit \Sigma}^{(2)}({\mib k},{\mit \omega}+{\rm i} {\mit \eta})}
          {Z({\mib k},{\mit \omega})},
\nonumber
\\
      {\mit \omega} Z({\mib k},{\mit \omega}) &=& {\mit \omega}
           - \frac{1}{2} \left[
                 {\mit \Sigma}^{(1)}({\mib k},{\mit \omega}+{\rm i} {\mit \eta})
               \right.
\nonumber
\\   && ~~~~~~~~~~ \left.
                -{\mit \Sigma}^{(1)}({\mib k},-{\mit \omega}-{\rm i} {\mit \eta})
             \right],
\label{eq:1}
\end{eqnarray}
${\mit \Sigma}^{(1)}$ and ${\mit \Sigma}^{(2)}$ being the normal and the 
anomalous self-energies, respectively. The symmetry of the gap function is 
clearly of ${\rm A}_2$ or $(x^2-y^2)$-type and its amplitude develops more 
rapidly than in the standard BCS model below $T_{\rm c}$. Figure 5 
shows the renormalized density of states at various temperatures, indicating 
how it is influenced by the d-wave gap formation with lowering temperature. As 
is expected from Fig. 4 the energy gap in the density of states develops 
rapidly with decreasing temperature, approaching close enough to a limiting 
result at around $T_{\rm c}/2$.
\begin{figure}[t]
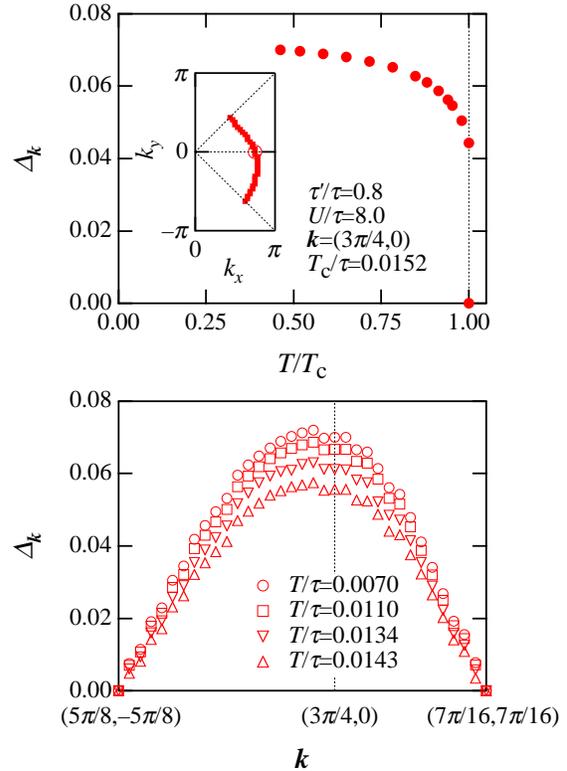

  \begin{center}
    \psbox[size=0.7#1]{fig4.eps}
  \end{center}
  \caption{
     Calculated gap function. (a) Temperature dependence of 
     ${\mit \Delta}_{\mibs k}$, with ${\mib k}$ indicated by an open circle
     on the Fermi line shown in the 
     inset. (b) Anisotropy or the wave vector-dependence of
     ${\mit \Delta}_{\mibs k}$, with ${\mib k}$ on the Fermi line 
     shown in the inset of (a). 
  }
\end{figure}
\begin{figure}[t]
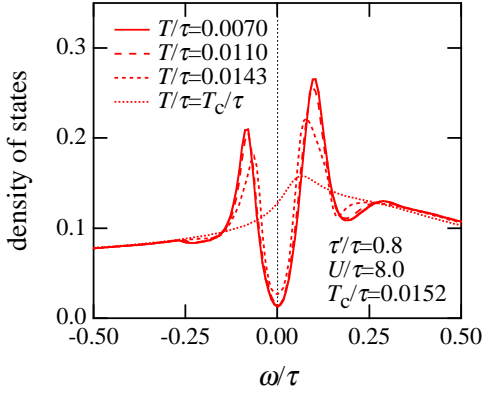

  \begin{center}
    \psbox[size=0.70#1]{fig5.eps}
  \end{center}
  \caption{
     Density of states at various temperatures.
  }
\end{figure}
\begin{figure}[t]
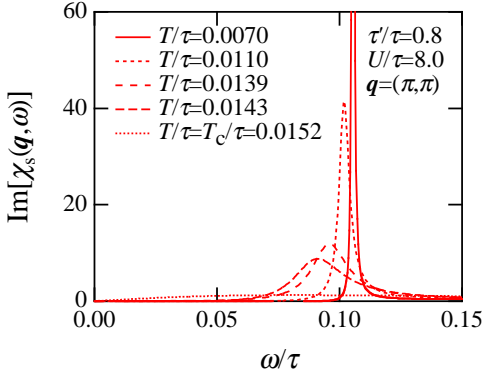

  \begin{center}
    \psbox[size=0.70#1]{fig6.eps}
  \end{center}
  \caption{
     Calculated imaginary part of the dynamical susceptibility 
     for ${\mib q}=({\mit \pi},{\mit \pi})$ at various temperatures.
  }
\end{figure}

The calculated dynamical susceptibilities well below $T_{\rm c}$ show strong 
resonance peaks around $({\mit \pi},{\mit \pi})$ and $({\mit \pi},-{\mit \pi})$, 
just as in the calculations for the high-$T_{\rm c}$ cuprates. Similarly to 
those in cuprates we interpret this resonance peak as a spin 
exciton,~\cite{rf:12,rf:17} a bound pair of an electron and a hole 
excited across the energy gap. Figure 6
shows the imaginary part of the dynamical spin susceptibility at 
${\mib q} = ({\mit \pi},{\mit \pi})$ at various temperatures. We see how the 
resonance peak develops from a broad spectrum in the normal state.  Figure 7
shows the dispersion and broadening of the spin excitons.  The resonance peak 
position ${\mit \omega}_{\mibs q}$ and  the width ${\mit \gamma}_{\mibs q}$ 
are given by
\begin{eqnarray}
     1 - U {\rm Re} {\overline {\mit \chi}}_{\rm s}
               ({\mib q}, {\mit \omega}_{\mibs q}) = 0,
\nonumber
\\
      {\mit \gamma}_{\mibs q} =
          \frac{{\rm Im} 
                      {\overline {\mit \chi}}_{\rm s} 
                                  ({\mib q}, {\mit \omega}_{\mibs q})}
                {\left\{
                \partial {\rm Re} {\overline {\mit \chi}}_{\rm s}
                                     ({\mib q},{\mit \omega})
                     / \partial {\mit \omega}
                \right\}_{{\mit \omega}={\mit \omega}_{\mibs q}}},
\label{eq:2}
\end{eqnarray}
with
\begin{eqnarray}
    {\overline {\mit \chi}}_{\rm s} 
                 \left( {\mib q} , {\rm i} {\mit \omega}_m \right) &=&
      - \frac{T}{N_0} \sum_{{\mibs k} , n} \left[
         G \left( {\mib k}+{\mib q} , 
                     {\rm i} {\mit \omega}_n + {\rm i} {\mit \omega}_m \right) 
         G \left( {\mib k} , {\rm i} {\mit \omega}_n \right) \right.
\nonumber  \\
      && \left.
       + F \left( {\mib k}+{\mib q} , 
                     {\rm i} {\mit \omega}_n + {\rm i} {\mit \omega}_m \right) 
         F \left( {\mib k} , {\rm i} {\mit \omega}_n \right) \right],
\label{eq:3}
\end{eqnarray}
where ${\mit \omega}_m$ and ${\mit \omega}_n$ are the Bose and Fermi Matsubara 
frequencies, respectively, and 
$G \left( {\mib k} , {\rm i} {\mit \omega}_n \right)$ and 
$F \left( {\mib k} , {\rm i} {\mit \omega}_n \right)$ are the renormalized 
normal and anomalous Green's functions, respectively. The resonance peak 
appears only in limited regions of the ${\mib q}$-space around 
$({\mit \pi},{\mit \pi})$ and $({\mit \pi},-{\mit \pi})$.  We show in Fig. 8
the temperature variation of the ${\mib q}$-integrated intensity spectrum. 
It is desirable to have the corresponding neutron inelastic scattering 
experiments on single crystals in future although such an experiment,
with a deuterized sample,  
does not seem to be very easy.
\begin{figure}[t]
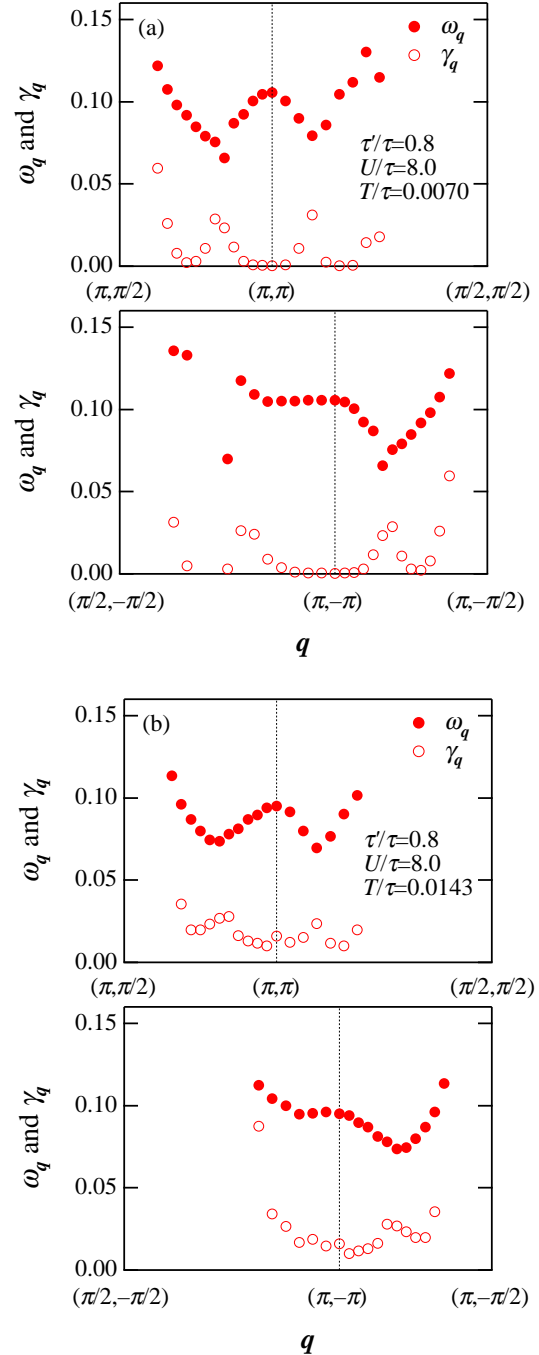

  \begin{center}
    \psbox[size=0.70#1]{fig7a.eps}
  \vspace*{5mm}
    \psbox[size=0.70#1]{fig7b.eps}
  \end{center}
  \caption{
     Dispersion and broadening of the resonance peak or the 
     spin excitons. (a) $T/{\mit \tau}= 0.0070$, (b) $T/{\mit \tau}= 0.0143$.
  }
\end{figure}
\begin{figure}[t]
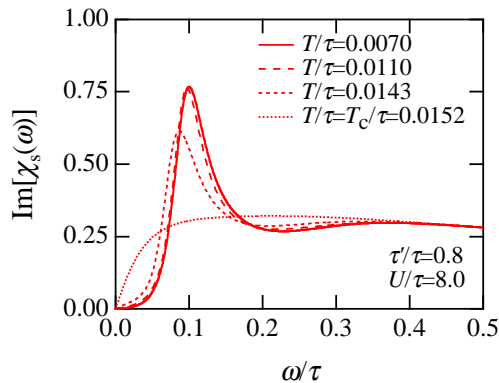

  \begin{center}
    \psbox[size=0.70#1]{fig8.eps}
  \end{center}
  \caption{
     ${\mib q}$-integrated spectrum of the imaginary part of 
     the dynamical susceptibility.
  }
\end{figure}
\begin{figure}[t]
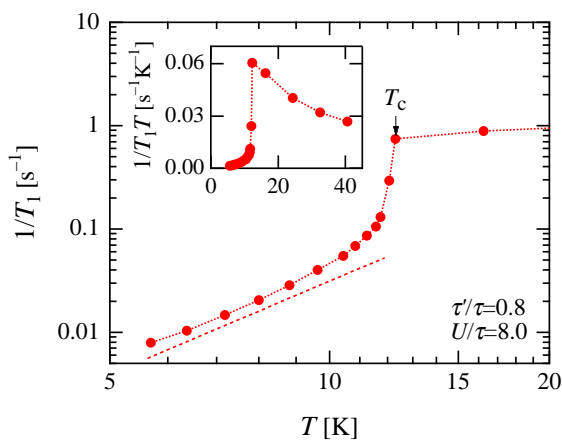

  \begin{center}
    \psbox[size=0.65#1]{fig9.eps}
  \end{center}
  \caption{
     Calculated nuclear spin-lattice relaxation rate $1/T_1$ 
     (logarithmic plots). The dashed line indicates a $T^3$ behavior. 
     Inset shows $1/ T_1 T$ (normal plots).
  }
\end{figure}

The nuclear spin-lattice relaxation rate can be calculated from the following 
general formula~\cite{rf:28} using the FLEX dynamical susceptibilities:
\begin{eqnarray}
    \frac{1}{T_1} = 
      \frac{{{\mit \gamma}_{N}}^2 T}{N_0} \sum_{\mibs q} 
                 \left| A_{\mibs q} \right|^2 
            \frac{{\rm Im} {\mit \chi}^{-+} ({\mib q}, {\mit \omega}_{0})}
                 {{\mit \omega}_{0}} ,
\label{eq:4}
\end{eqnarray}
where ${\mit \gamma}_{N}$ is the nuclear gyro-magnetic ratio, $A_{\mibs q}$ the 
Fourier ${\mib q}$-component of the hyperfine coupling constant, and 
${\mit \omega}_{0}$ the resonance frequency. We may neglect the 
${\mib q}$-dependence of the hyperfine coupling constant which is anisotropic 
at the site of $^{13}{\rm C}$. According to ref. 8 the principal values 
parallel and perpendicular to the direction of the 
$^{13}{\rm C}$=${^{13}{\rm C}}$ bond are $a+2B$ and $a-B$, respectively, 
with $a = 1.3 {\rm kOe}/{\mit \mu}_{\rm B}$ and  
$B = 2.1 {\rm kOe}/{\mit \mu}_{\rm B}$ for one spin per molecule; thus these 
values should be divided by 2 for one spin per dimer. For experimental results 
under the magnetic field parallel to the conducting layer 
we may approximately replace $\left| A_{\mibs q} \right|^2$ 
in eq. (4) with $[(a+2B)^2 + (a-B)^2]/8$, 
since the 
${\rm C}$=${\rm C}$ bond direction is nearly perpendicular to the layer. 
The result of calculation using this value 
for the coupling constant is shown in Fig. 9.
The $T^3$ dependence of the relaxation rate below $T_{\rm c}$ as observed 
experimentally is well reproduced.~\cite{rf:22,rf:23,rf:24} 
The order of magnitude seems also reasonable, 
in view of the approximate nature of the calculation. 
The observed values of $1/ T_1 T$ above $T_{\rm c}$ has a peak at 
$T = T^{\ast} \sim 50 {\rm K}$. Above this temperature $T^{\ast}$ the system 
shows insulating behavior while the system is metallic below $T^{\ast}$. Thus 
above $T^{\ast}$ we had better use the Heisenberg model with the Anderson 
superexchange.

For $T^{\ast} > T > T_{\rm c}$  the observed values for  $1/ T_1 T$ decrease 
with decreasing temperature, reminiscent of the pseudo-spin gap behaviors in 
high-$T_{\rm c}$ cuprates. This behavior is hard to understand within the 
present model and approximation. One possible explanation may be given by 
considering a coexistence of spin density wave (SDW) fluctuations and a charge 
density wave (CDW) or its slow fluctuations, although further studies are 
necessary before drawing any conclusion. 

In summary, we have studied the spin fluctuation-induced superconducting 
state of quasi-two dimensional $\kappa$-$({\rm BEDT}$-${\rm TTF})_2 X$@
compounds by using a 
half-filled triangular Hubbard model within the FLEX approximation,
extending the previous study above $T_{\rm c}$. The energy gap of 
$(x^2-y^2)$-type develops below $T_{\rm c}$ more rapidly than in the BCS model. 
Resonance peaks in the dynamical susceptibility are predicted in limited regions
of the ${\mib q}$-space around $({\mit \pi},{\mit \pi})$ and 
$({\mit \pi},-{\mit \pi})$. The observed $T^3$ behavior of the nuclear 
spin-lattice relaxation rate $1/T_1$ is well reproduced with 
a reasonable order of 
magnitude. We have also predicted the doping concentration dependence of 
$T_{\rm c}$ in these compounds. Calculations of the SC and AF instability 
points in the $U/{\mit \tau}$ vs. ${\mit \tau}^{\prime} / {\mit \tau}$ plane
show that the former is favorable only for 
${\mit \tau}^{\prime} / {\mit \tau}$ larger than $\sim 0.3$, 
indicating that an antiferromagnetic 
instability dose not always accompany the d-wave superconductivity.

Finally we would like to emphasize that the present mechanism seems to be the 
only available mechanism to describe the anisotropic superconductivity in 
quasi-two dimensional $\kappa$-type BEDT-TTF compounds.~\cite{rf:21} 
Theoretical results seem to be generally consistent with available experimental 
results except for the pseudo-spin gap behavior of $1/T_1$ above
$T_{\rm c}$.~\cite{rf:29} 
Further experimental and theoretical investigations are desired to confirm 
this mechanism.  

We would like to thank Dr. S. Nakamura and Dr. T. Takimoto 
for helpful discussions.

%\newpage
%{\large {\bf Figure Captions}}

%\begin{description}
%\item[Fig. 1.~~] The model unit cell and the transfer integrals.
%\item[Fig. 2.~~] Instability points of the paramagnetic state against 
%     superconductivity and antiferromagnetism. Inset shows the results 
%     of an improved calculation for $T_{\rm c}/{\mit \tau}$ vs. $U/{\mit \tau}$.
%\item[Fig. 3.~~] Electron number (doping concentration) dependence of 
%     $T_{\rm c}/{\mit \tau}$ for ${\mit \tau}^{\prime} / {\mit \tau} = 0.8$.
%\item[Fig. 4.~~] Calculated gap function. (a) temperature dependence of 
%     ${\mit \Delta}_{\mibs k}$, with ${\mibs k}$ on the Fermi line shown in the 
%     inset. (b) anisotropy or the wave vector- dependence of
%     ${\mit \Delta}_{\mibs k}$, with ${\mibs k}$ on the Fermi line 
%     shown in the inset of (a). 
%\item[Fig. 5.~~] Density of states at various temperatures.
%\item[Fig. 6.~~] Calculated imaginary part of the dynamical susceptibility 
%     for ${\mib q}=({\mit \pi},{\mit \pi})$ at various temperatures. 
%\item[Fig. 7.~~] Dispersion and broadening of the resonance peak or the 
%     spin excitons. (a) $T/{\mit \tau}= 0.0070$ (b) $T/{\mit \tau}= 0.0139$.
%\item[Fig. 8.~~] ${\mib q}$-integrated spectrum of the imaginary part of 
%     the dynamical susceptibility.
%\item[Fig. 9.~~] Calculated nuclear spin-lattice relaxation rate $1/T_1$ 
%     (logarithmic plots). The dashed line indicates a $T^3$ behavior. 
%     Inset shows $1/ T_1 T$ (normal plots).
%\end{description}


\begin{thebibliography}{99}
\bibitem{rf:27} In solving the eigen value problem to obtain $T_{\rm c}$,
     we took account of all the $( {\mib k} , {\rm i} {\mit \omega}_n )$ points
     by using the Lanczos algorithm, while 1000 important points were taken into
     account in the previous calculation in I. New values of $T_{\rm c}$
     are somewhat larger than those in I.
\bibitem{rf:21} For example, any theories based on the $t$-$J$ model can 
     clearly be excluded since the half-filled electrons in these systems are 
     in an intermediate coupling regime where an expansion in ${\mit \tau}/U$ 
     is inapplicable.
\end{thebibliography}
\end{document}